\documentclass[aps, showpacs, draft, eqsecnum, superscriptaddress]{revtex4}

\usepackage{epsf}  
\usepackage[final]{graphicx}  
\usepackage{amsmath,amssymb} 

\usepackage{mathrsfs} 
\usepackage{psfrag}

\newcommand{\bigO}{\mathcal O}
\newcommand{\pd}[2]{\frac{\partial #1}{\partial #2}}
\newcommand{\Vr}{\frac{V}{r}}
\def\NP{c_{\scriptscriptstyle\text{NP}}}
\newcommand{\Scri}{{\mathscr I}}
\newcommand{\Scrh}{{\mathscr H}}
\newcommand{\mEXT}{m_{\scriptscriptstyle\text{EXT}}}
\newcommand{\mSSH}{m_{\scriptscriptstyle\text{SSH}}}
\newcommand{\SSH}{{\scriptscriptstyle\text{SSH}}}
\newcommand{\mB}{m_{\scriptscriptstyle\text{B}}}
\newcommand{\mBH}{m_{\scriptscriptstyle\text{BH}}}
\newcommand{\uB}{u_{\scriptscriptstyle\text{B}}}
\newcommand{\tauB}{\tau_{\scriptscriptstyle\text{B}}}

\DeclareFontFamily{OT1}{rsfs}{}
\DeclareFontShape{OT1}{rsfs}{m}{n}{ <-7> rsfs5 <7-10> rsfs7 <10->
rsfs10}{} \DeclareMathAlphabet{\mycal}{OT1}{rsfs}{m}{n}

\begin{document}

\title{News from Critical Collapse: Bondi Mass, Tails and Quasinormal Modes}

\author{Michael P\"urrer}
\affiliation{
        Institut f\"ur Theoretische Physik,
        Universit\"at Wien,
        1090 Wien, Austria}

\author{Sascha Husa}
\affiliation{
        Max-Planck-Institut f\"ur Gravitationsphysik,
        Albert-Einstein-Institut,
        14476 Golm, Germany}

\affiliation{
        Departament de Fisica,
        Universitat de les Illes Balears,
        Ctra de Valldemossa km 7,5,
        07071 Palma de Mallorca, Spain}

\author{Peter C. Aichelburg}
\affiliation{
        Institut f\"ur Theoretische Physik,
        Universit\"at Wien,
        1090 Wien, Austria}

\date{\today}

\begin{abstract}
We discuss critical gravitational collapse on the threshold 
of apparent horizon formation as a model both for the discussion of
global aspects of critical collapse and for numerical studies in a
compactified context.
For our matter model we choose a self-gravitating massless
scalar field in spherical symmetry, which has been studied extensively
in the critical collapse literature.
Our evolution system is based on Bondi coordinates, the mass function is
used as an evolution variable to ensure regularity at null
infinity. We compute radiation quantities like the Bondi mass and news 
function and find that they reflect the discretely self-similar (DSS) 
behavior. Surprisingly, the
period of radiation at null infinity is related to the formal result
for the leading quasi-normal mode of a black hole with rapidly
decreasing mass. Furthermore, our investigations
shed some light on global versus local issues in critical
collapse, and the validity and usefulness of the
concept of null infinity when predicting detector signals.
\end{abstract}

\maketitle

\section{Introduction}

In this paper we present a numerical study of 
scalar field critical collapse on the threshold of singularity formation.
We consider Einstein's equations with a minimally coupled massless scalar field
$\phi$:
\begin{eqnarray}
\label{Einstein}
G_{ab} &=& 8 \pi T_{ab} = \nabla_a \phi \nabla_b \phi - \frac{1}{2} g_{ab} \nabla_c \phi \nabla^c \phi,\\
\label{WaveEq}
\Box_g \phi &=& 0,
\end{eqnarray}
We restrict ourselves to the case of spherical symmetry, which
has been studied extensively with numerical and analytical techniques
since the discovery of critical collapse by Choptuik 
\cite{Choptuik92,Choptuik93}.
We extend previous investigations by focussing on global aspects
of this problem, and use a compactified evolution scheme which includes
null infinity on our numerical grid.
The motivation is twofold: First, we want to simplify the discussion and
improve the understanding of local versus global issues in critical collapse.
In particular, we try to explain questions like:
What is the role of asymptotic flatness for critical collapse 
(e.g. the critical solution, the ``Choptuon'' is self-similar, 
and thus not asymptotically flat)?
How would hypothetical detectors of radiation observe the dynamics 
close to criticality?
How can we understand the way null infinity approximates observers at large 
distances in this simple but nontrivial setup?
The second motivation is to test numerical algorithms which are based on
compactification methods in a situation that is very demanding on accuracy.
We will argue that at least in the model considered here, global methods
do not cause a significant penalty in accuracy, but simplify the
interpretation of certain results.

Critical phenomena in gravitational collapse have been
originally discovered in the pioneering numerical investigations of
Choptuik \cite{Choptuik92,Choptuik93}. 
He studied a massless scalar field coupled
to gravity with sophisticated numerical techniques that allowed him to
analyze the transition in the space of initial data
between dispersion to infinity and the formation of a black hole.
It turned out, that black holes of arbitrarily small mass can be created, and
that the critical solutions approach a discretely self-similar solution,
called the ``Choptuon''. Both  the Choptuon and the scaling law for the black
hole mass are universal for arbitrary families of initial data.
Hamad\'e and Stewart \cite{Stewart96} have found numerical evidence, that the
critical solution contains a naked singularity which can be seen at future
null infinity.
Similar critical solutions -- exhibiting (continuous or discrete)
self-similarity --
have also been found for several other types of matter fields, and have been
constructed directly in several cases \cite{Gundlach95,Gundlach97f,Gundlach96a,Lechner2001,LechnerPhD}.
The problem has also been studied extensively from an analytic point of view by
Christodoulou \cite{Christodoulou86,Christodoulou87,Christodoulou91,Christodoulou94}, 
in particular he could show that the space
of regular initial data that lead to naked singularities has measure zero
\cite{Christodoulou99}.

In the current work, we refer to critical collapse phenomena as
``critical collapse at the threshold of apparent horizon formation'' to avoid
possible misunderstandings, since critical collapse is essentially a quasilocal
phenomenon and the standard definition of black holes is based on global
concepts (see textbooks like e.g. \cite{Wald84}).
Also, this term emphasizes the relation of these phenomena to other 
areas in nonlinear partial differential equations (PDEs), where related phenomena occur, 
but the concept of black holes is absent.

Critical behavior of the kind originally found by Choptuik is usually referred
to as type II, because of its formal correspondence with type II phase
transitions of statistical physics.
A different type of critical solutions at the threshold of black hole
formation, corresponding to type I phase transitions, is provided by unstable
static configurations  -- like those found by Bartnik and McKinnon \cite{Bartnik88}.

Linear perturbation calculations of such critical solutions revealed exactly
one unstable mode, which confirmed their interpretation
as intermediate attractors in the language of dynamical systems.
Critical phenomena in general
relativity are reviewed in \cite{Gundlach97d, Bizon96},
including discussions in
terms of phase transitions and renormalization group
techniques familiar from statistical physics.

A massless scalar field in spherical symmetry 
exhibits type II critical collapse (there are no regular stationary or 
time-periodic solutions).
Type II critical solutions have been found to exhibit continuous or discrete
self-similarity in the past lightcone of the singularity.
In our case, the critical solution is known to be
discretely self-similar (DSS), and has been constructed directly as an
eigenvalue problem \cite{Gundlach95}.

A spacetime is said to be DSS 
\cite{Gundlach00a}
if it admits a discrete diffeomorphism $\Phi_\Delta$ which leaves the 
metric invariant up to a constant scale factor: 
\begin{equation}\label{DSS-def}
\left(\Phi^*_\Delta\right)^n g = e^{2 n \Delta} g,
\end{equation}
where $\Delta$ is a dimensionless real constant and $n \in \mathbb{N}$.

We choose scalar field critical collapse in spherical
symmetry for several reasons: the model is very well studied and we can 
compare with a large amount of previous numerical and analytical results. 
Furthermore, the model is also very demanding:
The value of the echoing period in the DSS critical solution 
is $\Delta \simeq 3.44$, which is quite larger compared to
many other models. Note that larger values of $\Delta$ make it 
more difficult to resolve a large number of echos.

Our numerical method is based on a characteristic initial value problem, i.e.
we foliate spacetime by 
null cones. This allows for a very efficient evolution system and
simplifies the study of the causal structure of the solutions. 
In spherical symmetry, caustics are restricted to the center of symmetry,
so we do not have to deal with
the dynamical appearance of caustics, which causes potential 
problems for characteristic initial value problems in higher dimensions.
Our numerical approach
mixes techniques from previous work of Garfinkle \cite{Garfinkle95}
and the Pittsburgh group \cite{Gomez92a,Winicour98}, in particular we follow Garfinkle
in moving along ingoing null geodesics to utilize
gravitational focusing for increasing resolution in the region of large 
curvature. 
Furthermore, compactification methods are well studied and relatively 
straightforward to implement in characteristic codes. 

\begin{figure}
\centering
\begin{psfrags}
\psfrag{r=0}[][m]{$\scriptstyle r=0$}
\psfrag{i+}[][m]{$\scriptstyle i^+$}
\psfrag{i-}[][m]{$\scriptstyle i^-$}
\psfrag{i0}[][m]{$\scriptstyle i^0$}
\psfrag{scri+}[][m]{$\scriptstyle \Scri^+$}
\psfrag{scri-}[][m]{$\scriptstyle \Scri^-$}
\psfrag{H+}[][m]{$\scriptstyle \Scrh^+$}
\includegraphics[width=.3\textwidth]{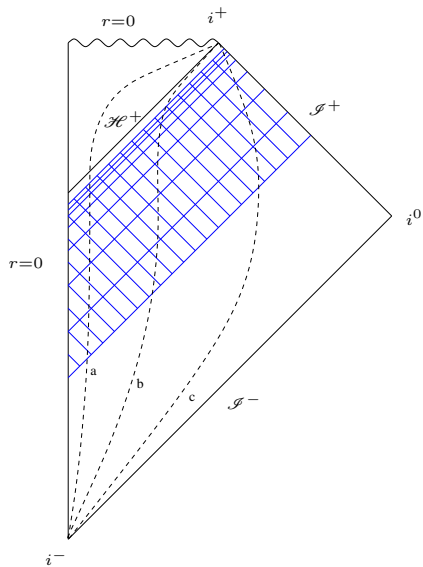}
\end{psfrags}
\caption{\label{spacetime}A Penrose diagram of a typical collapse spacetime. 
Shown is our numerical null grid which extends to future null infinity $\Scri^+$.
The grid consists of the null slices $u=const$ and ingoing radial null geodesics $v=const$.
Evolution slows down in the vicinity of the future event horizon $\Scrh^+$.
We also indicate lines (a) $r = const < 2M_f$, (b) $r = const = 2M_f$
and (c) $r = const > 2M_f$, where $M_f$ is the final black hole mass.
}
\end{figure}

An important aspect of our compactified characteristic evolution scheme is that
at late times our null slices asymptotically 
approach the event horizon, see Fig. \ref{spacetime}.
Essentially this is because our coordinates can not penetrate a 
dynamical horizon \cite{Ashtekar02,Ashtekar03,Ashtekar04,Hayward94}
(they become singular at a marginally trapped surface, e.g.
at an apparent horizon),
which is spacelike if any matter or radiation falls through it and
null otherwise \cite{Hawking73a,Ashtekar04}.
Note that the dynamical horizon is contained inside of the event horizon,
and the outermost dynamical horizon approaches (or coincides with) the
event horizon at late times, assuming cosmic censorship holds.
This fact makes our approach in some sense complementary to previous critical
collapse studies, which were not adapted
to the asymptotic regime. In this paper we focus on those aspects of critical
collapse which are associated with global structure, and in particular the 
phenomenology seen by asymptotic observers. We will only make brief remarks about 
other well-studied aspects such as mass scaling and universality.

The paper is organized as follows: in Sec. \ref{sec:model} we give a brief
introduction to scalar field critical collapse in spherical
symmetry, and discuss our geometric setup, which is based on Bondi-type coordinates
and a compactification scheme which introduces the Misner-Sharp mass function as an independent 
evolution variable, which renders our evolution system regular at null infinity.
Our numerical algorithms are presented in Sec. \ref{sec:algorithm}.
In Sec. \ref{sec:results} we give a detailed discussion of our numerical
results, which include
the study of power law tails, and compare the radiation signal at null infinity with a heuristic
estimate based on self-similar scaling and quasi-normal mode frequencies.
Our results and conclusions are summarized in Sec. \ref{sec:discussion}.

\section{The Continuum Problem}\label{sec:model}

\subsection{Geometric Setup}\label{sec:geometry}

We assume spherical symmetry with a regular center.
We introduce a Bondi coordinate system $\{u,r,\theta,\varphi\}$ on spacetime
based upon outgoing null hypersurfaces $u = \text{constant}$, with the line element
\begin{equation}
ds^2 = -e^{2\beta(u,r)} du \Bigl(\frac{V(u,r)}{r} du + 2 dr \Bigr) + r^2 (d\theta^2 + \sin^2\theta d\varphi^2),
\end{equation}
and assume that spacetime admits a regular center $r = 0$.
Smothness of the metric requires the metric functions $\beta$ and $V/r$
to be smooth functions on the spacetime manifold. Note that $r$ is not a 
smooth function in this sense.
Consequently, at fixed retarded time $u_0$, $\beta$ and $V$ behave as
\begin{equation}\label{eq:beta-V-smoothness}
\begin{split}
\beta(u_0,r) &= \bigO(r^2),\\
V(u_0,r) &= r + \bigO(r^3),
\end{split}
\end{equation}
where the gauge has been fixed such that the family of outgoing null cones
emanating from the center is parametrized by the proper time $u$ at the center.

Following earlier work \cite{Gomez92a,Gomez92,Gomez92b}, we write the curved space wave 
equation (\ref{WaveEq}), $\Box_g \phi = 0$, in terms of a rescaled field
$\psi = \phi \, r$ as 
\begin{equation}\label{wave-eq-int}
\Box_h \psi - \Bigl(\Vr\Bigr)_{,r} \frac{e^{-2\beta}\psi}{r} = 0,
\end{equation}
where $\Box_h$ is the 2 dimensional wave operator in the $(u,r)$ submanifold.
This ansatz factors out the known falloff of $\phi$ at large distances.
The evolution system is completed by 
two hypersurface equations for the metric functions (which follow from the
$(r,r)$ and $(u,r)$ components of Einstein's equations, respectively)
\begin{equation}\label{hyper-eq}
\begin{split}
\beta_{,r} &= 2\pi r \left( \phi_{,r} \right)^2,\\
V_{,r} &= e^{2\beta}.
\end{split}
\end{equation}

In spherical symmetry, there exists a well defined notion of quasilocal energy,
the Misner-Sharp mass-function \cite{Misner-Sharp-1964}:
\begin{equation}
m(u,r) = \frac{r}{2} \left[ 1 - \Vr e^{-2\beta} \right].
\end{equation}
Note that $m/r$ is a smooth function. The 
Misner-Sharp mass measures the energy content of a sphere fo radius $r$ and 
reduces the Arnowitt-Deser-Misner (ADM) and Bondi masses in the appropriate limits.

In adapted coordinates 
\begin{eqnarray}
  \label{eq:tau-z-def}
\tau &=& -\ln \frac{u^* - u}{u^*},\\
z    &=& \frac{r}{(u^* - u)\zeta(\tau)} = \frac{r e^\tau}{\zeta(\tau) u^*},\label{eq:z-def} 
\end{eqnarray}
where $u^*$ is a real number which denotes 
the accumulation time of DSS and $\zeta(\tau + \frac{\Delta}{2}) = \zeta(\tau)$, we have
\begin{equation}
f^*(\tau + n\Delta, z) = f^*(\tau, z),
\end{equation}
where $f^*$ denotes $\phi$ or the metric functions $\beta$, $\Vr$, which are 
defined in section \ref{sec:geometry}. 
In addition $\phi$ also satisfies
\begin{equation}
\left(\Phi^*_{\Delta/2}\right)^n \phi = \left(-1\right)^n \phi,
\end{equation}
so that fields even in $\phi$, such as $\beta, \Vr, m$, are periodic in $\tau$
with period $\Delta/2$.

\subsection{Compactification}

Compactification is rather simple for characteristic codes, and has
been used extensively in the  characteristic approach to numerical
relativity \cite{Gomez92a,Gomez92,Gomez92b}.
Before introducing compactification we want to say a few words about
asymptotic series expansions.
Assuming initial data that are smooth at $\Scri^+$, one can
expand the massless scalar field $\phi$ in powers of $1/r$ near $\Scri^+$ 
\begin{equation}\label{phi_asm}
\phi(u,r) = \frac{c(u)}{r} + \frac{\NP}{r^2} + O(r^{-3}).
\end{equation}
The coefficient $\NP$ of the $1/r^2$- term 
in the expansion is a Newman-Penrose constant \cite{Newman68}
of the scalar field.
Inserting the expansion \eqref{phi_asm} into the hypersurface equations 
\eqref{hyper-eq} yields
\begin{equation}
\beta(u,r) = H(u) - \frac{\pi c^2(u)}{r^2} + O(r^{-3}),
\end{equation}
and
\begin{equation}
V(u,r) = e^{2H(u)} \left(r - 2M(u) + \frac{\pi c^2(u)}{r}\right) + O(r^{-2}),
\end{equation}
where integration constants $H(u)$ and $M(u)$ have been introduced.
$H(u)$ indicates redshift since Bondi time $\uB$ is related to proper 
time at the center via the relation
\begin{equation}\label{eq:def_BondiTime}
\frac{d\uB}{du} = e^{2H(u)}.
\end{equation}
$M(u) = \lim_{r\to\infty} m(u,r)|_{u=const}$ is the Bondi mass which
is in general not conserved.
The Bondi mass-loss equation (derived from the ${uu}$ component 
of Einstein's equations) states
\begin{equation}\label{eq:mass-loss}
\frac{dM}{d\uB} = -4\pi N(\uB)^2,
\end{equation}
where the news-function is defined as
\begin{equation}
N(\uB) = \frac{dc}{d\uB}.
\end{equation}

In order to write down a compactified evolution system,
we introduce a compactified radial coordinate
\begin{equation}
x := \frac{r}{1+r},
\end{equation}
so that points at $\Scri^+$ are automatically included in the grid at $x=1$.
Our aim is to regularize our equations at null infinity, the coordinate 
singularity at the regular center can be dealt with in a straightforward 
manner, see \cite{Husa2000b}.

A naive approach of rewriting the hypersurface equations in terms 
of the $x$-coordinate leads to a singular equation for the quantity $V$
\begin{equation}
\begin{split}
\beta_{,x} &= 2\pi x (1-x) (\phi_{,x})^2,\\
V_{,x} &= \frac{e^{2\beta}}{(1-x)^2}.
\end{split}
\end{equation}
To obtain a regular system of evolution equations, we eliminate $V$ by the
Misner-Sharp mass-function $m(u,r)$.
The set of hypersurface equations then becomes:
\begin{equation}
\begin{split}
\frac{dm}{dx} &= 2\pi x^2 \left[ 1 - \frac{2(1-x)}{x} m \right] \left( \phi_{,x} \right)^2,\\
\beta_{,x} &= 2\pi x (1-x) (\phi_{,x})^2.
\end{split}
\end{equation}
Note that these equations are completely regular.

We choose our gridpoints to freely fall along ingoing radial null geodesics 
$x(u)$ which fulfill
\begin{equation}
\frac{dx}{du} = -\frac{1}{2} (1-x)^2 e^{2\beta} \left(1 - 2m \frac{1-x}{x} \right).
\end{equation}
Note that the term $m \frac{1-x}{x} = m/r$ does not cause problems because of 
the smoothness of the metric at the regular center \eqref{eq:beta-V-smoothness}.
In section \ref{sec:amr} we will argue that this choice is crucial to resolve DSS
phenomena.

The vanishing of the outgoing null expansion
\begin{equation}
\Theta_+ = \frac{2}{r}e^{-2\beta}
\end{equation}
on some 2-sphere $r=\text{constant}$
means that this sphere is marginally outer trapped. 
Since this requires $\beta$ to diverge, we cannot penetrate apparent 
horizons in Bondi coordinates.

\section{Numerical Algorithm}\label{sec:algorithm}

\subsection{Evolution}\label{sec:evolution}

\begin{figure}
\centering
\begin{psfrags}
 \psfrag{u0}[lb][m]{$\scriptstyle u_0$}
 \psfrag{u1}[lb][m]{$\scriptstyle u_0 + \Delta u$}
 \psfrag{r0}[lb][m]{$\scriptstyle r=0$}
 \psfrag{N}[][r]{$\scriptstyle N$}
 \psfrag{S}[][r]{$\scriptstyle S$}
 \psfrag{W}[][r]{$\scriptstyle W$}
 \psfrag{E}[][r]{$\scriptstyle E$}
\includegraphics[width=.3\textwidth]{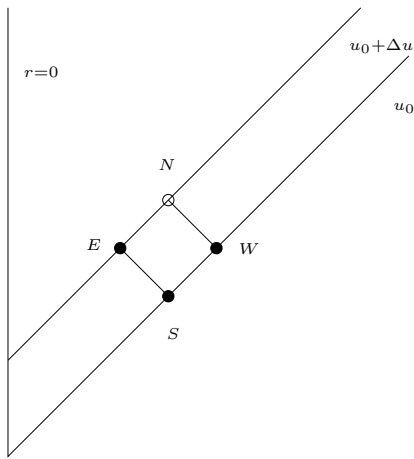}
\end{psfrags}
\caption{\label{NSWE}A representative null-paralellogram in the numerical grid made up by
two $u=const$ surfaces and two ingoing null geodesics $v=const$.}
\end{figure}

We choose to evolve the scalar field $\phi$ by the diamond scheme 
outward marching algorithm \cite{Winicour98}
which can be obtained by integrating \eqref{wave-eq-int} over a null parallelogram $\Sigma$
(see figure \ref{NSWE}):
\begin{equation}\label{eq:phi-evo-continuum}
  \int_\Sigma \Box_h \psi = \int_\Sigma du dr \Bigl(\Vr\Bigr)_{,r} \frac{\psi}{r}.
\end{equation}

Since any 2 dimensional metric is conformally flat and the conformal weights of
$\Box_h$ and the surface area element $d^2 x \sqrt{-h}$ cancel, 
the integral $\int_\Sigma \Box_h \psi$ is equal to the flat space result 
$-2 \int_\Sigma du dv \, \psi_{,uv}$ and we obtain
\begin{equation}\label{eq:phi-evo}
  \psi_N = \psi_W + \psi_E - \psi_S - \frac{1}{2} \int_\Sigma du dr \Bigl(\Vr\Bigr)_{,r} \frac{\psi}{r}.
\end{equation}
In terms of the $x$-coordinate this scheme becomes
\begin{equation}
\psi_N = \psi_W + \psi_E - \psi_S - \int_\Sigma du dx  \, e^{2\beta} \, m \, \psi \, \frac{1-x}{x^3}.
\end{equation}

The gauge and regularity conditions for the metric functions and 
the massless scalar field at the origin of spherical symmetry become:
\begin{equation}
\begin{split}
\beta(u,x) &= \mathcal O (x^2),\\
m(u,x)     &= \mathcal O (x^3),\\
\psi(u,x)  &= \mathcal O (x). 
\end{split}
\end{equation}

Our code is based on the ``DICE''
(Diamond Integral Characteristic Evolution) code, which has been documented in
\cite{Husa2000b} (there particular emphasis is given to detailed
convergence tests).

\begin{figure}
\begin{center}
\begin{psfrags}
\psfrag{x}[][m][1][0]{$x$}
\psfrag{abs(Euur,u=2.033)}[][m][0.8][0]{$|\mathbf{E_{uur}}(u=2.033,x)|$}
\includegraphics[width=.5\textwidth]{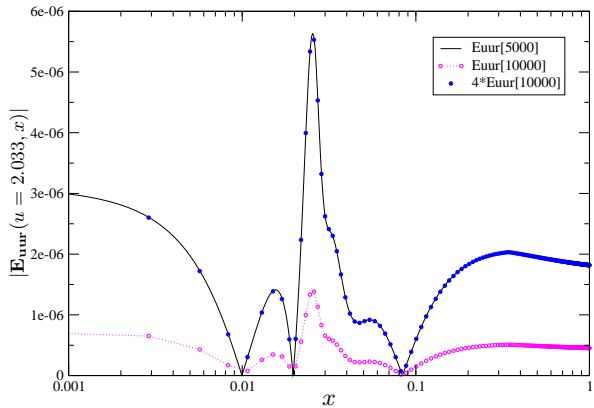}
\end{psfrags}
\caption{\label{fig:convergence}The convergence of the error diagnostic 
$\mathbf{E_{uur}}$ with increasing grid resolution for two near-critical evolutions. 
Evolution (1) uses 5000 gridpoints and $p=p^*[5000] + 10^{-10}$
and evolution (2) uses 10000 gridpoints and $p=p^*[10000] + 10^{-10}$. 
The data displayed as dots and circles have been sampled.}
\end{center}
\end{figure}

The code is globally second order accurate. 
Figure \ref{fig:convergence} shows a convergence test for near-critical
evolutions. 
We want to emphasize that the critical
value $p^*$ of the initial data parameter depends on the grid resolution.
This fact is essential when doing convergence tests for near critical
evolutions, as has been discussed in our previous paper \cite{Husa2000b}.
 
Timesteps are chosen adaptively via a condition
\begin{equation}
(V/r) \Delta u \le C \Delta r,
\end{equation}
where $C$ is a constant, on the order of unity. This restriction on 
$\Delta u$ is most severe at $\Scri^+$.
For various analysis purposes, gridfunctions can be evaluated at $x=const$ 
locations by determining the nearest 
gridpoint to the specified $x=const$ value and using cubic spline 
interpolation in a 5-gridpoint interval.

To monitor the accuracy of the code during runs we use components of the 
Einstein equations which are automatically satisfied if the evolution 
equations hold, such as the following linear combination of the $(u,u)$ 
and $(u,r)$ components 
\begin{equation}\label{eq:Euur1}
- E_{uur} \equiv r^2 \Bigl( G_{uu} - 8 \pi T_{uu} \Bigr) 
                        - r^2 (V/r) \Bigl( G_{ur} - 8 \pi T_{ur} \Bigr).
\end{equation}
This can be rewritten in the following form
\begin{equation}
E_{uur} = 2 e^{2\beta} \dot m + 8\pi \left[ \dot\psi^2 - e^{2\beta}\left(\frac{1}{1-x} - \frac{2m}{x}\right)
                                                         (1-x)^2 \dot\psi
                                                         \left(\psi_{,x} (1-x) - \frac{\psi}{x} \right)
                                     \right],
\end{equation}
where $\dot f = \pd{f}{u}|_x$.
Since this expression is a linear combination of tensor components,
we use a suitably normalized quantity ${\mathbf E_{uur}} = \frac{E_{uur}}{1 + E_{|uur|}}$, 
where $E_{|uur|}$ is the sum of the running maxima ($\max_{i\le j} f_i$ over a $u=const$-slice) 
of the absolute values of the individual terms of $E_{uur}$.
We must have $\mathbf{E_{uur}} \ll 1$ for our finite difference solution to be a good
approximation to the continuum solution.
Provided that there are no unexpected cancellations between the two terms in 
equation \eqref{eq:Euur1}, $\mathbf{E_{uur}}$ will be a reliable measure for the
accuracy of our code.
A convergence test for this quantity is shown in figure \ref{fig:convergence}.

\subsection{Mesh Refinment}\label{sec:amr}

\begin{figure}
\begin{center}
\begin{psfrags}
\includegraphics[angle=-90,width=.5\textwidth]{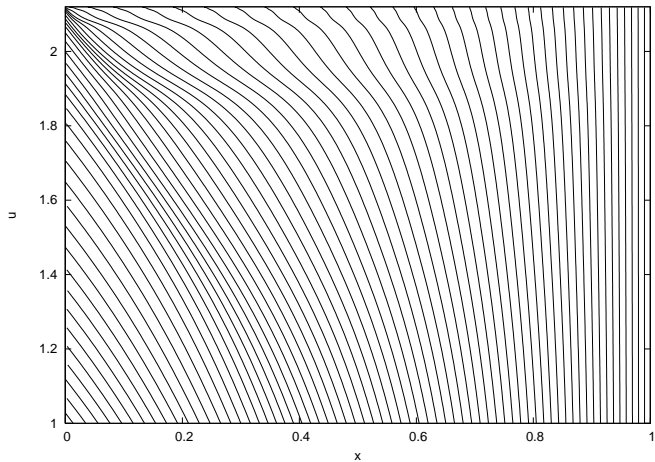}
\end{psfrags}
\caption{\label{fig:nullgeo}This figure shows the focussing of ingoing 
null-geodesics by gravity in the late stages of a slightly supercritical evolution.
The discretely self-similar dynamics causes the density of the geodesics to
increase in a periodic manner.
}
\end{center}
\end{figure}

Hamad\'e and Stewart \cite{Stewart96} have implemented full Berger-Oliger 
mesh refinement in double null coordinates (without compactification) to
achieve sufficient resolution to study critical collapse. Garfinke 
\cite{Garfinkle95} has shown, that this is not really necessary -- here
we follow his approach to increase resolution:
Most importantly, we choose our gridpoints to follow ingoing radial
nullgeodesics. This leads to a rapid loss of gridpoints in the early
phase of collapse, but to an accumulation of gridpoints in the region of strong
curvature for the late stages of critical collapse (see figure \ref{fig:nullgeo}). 
Furthermore, when half of the gridpoints have reached the origin we refine the grid
and thus obtain a very simple but effective form of mesh refinement
which is a crucial ingredient in the calculation of critical collapse spacetimes.
In previous work \cite{Husa2000b} we have also tuned our outermost gridpoint to
be located just outside of the self-similar horizon (SSH). Here we choose to go out 
all the way to null infinity.
The most effective approach in this situation would be to just refine the region inside
the SSH when half of the gridpoints in this region have reached the origin.
While this is straightforward to implement, we found the penalty on the 
resolution that the original condition (the loss of half of \emph{all} gridpoints) 
causes to be acceptable for the results presented here.

\section{Results}\label{sec:results}

\subsection{Identification of critical behavior}\label{sec:crit_behavior}

\begin{figure}
\centering
\includegraphics[clip,width=.5\textwidth]{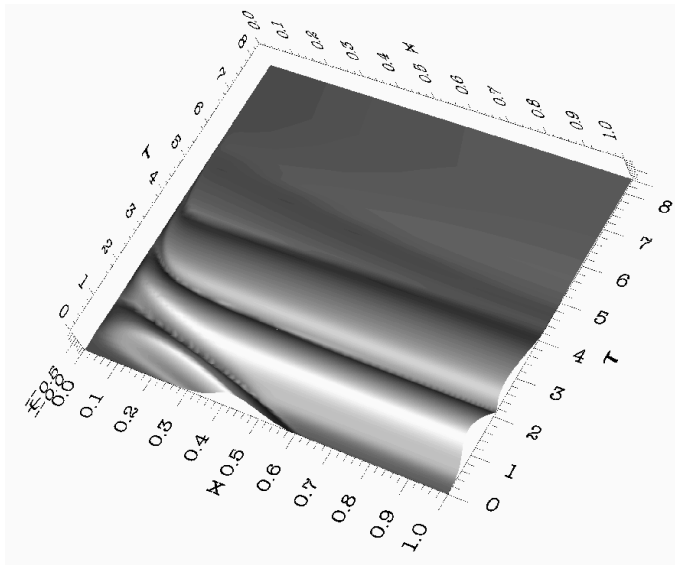}
\caption{\label{fig:psi-perspective}This figure shows a surface plot 
of $|\psi(\tau,x)|$ for 
a near-critical evolution.
When the initial Gaussian reaches the origin, it is  ``instantly'' (in retarded time $u$)
radiated to future null infinity $\Scri^+$ (located at $x=1$) by interfering nonlinearly with
the field that has not yet reached the origin. Once the evolution has come close to
the critical solution, the matter field $\psi(u,x) = \phi r$ decays exponentially;
further self-similar features are thus not visible in this plot.
}
\end{figure}

We consider 1-parameter families of initial data $\phi = \phi_p(u_0,x)$,
such that for small values of $p$ we have dispersion, while for large
values of $p$ we have black hole formation. 
It has been found numerically \cite{Choptuik92,Choptuik93} that, for any initial 
data family considered, the evolution of near-critical data approach a 
universal DSS critical solution.
For the present paper we already assume universality and restrict ourselves
to Gaussian-like initial data
\begin{equation}\label{Gaussian-ID}
\phi(u_0,x) = A \, r(x)^2 \exp \Bigl[ - \bigl(\frac{r(x) - r_0}{\sigma}\bigr)^2 \Bigr],
\end{equation}
where $r(x) = \frac{x}{1-x}$. This choice makes it easy to compare 
compactified to uncompactified evolutions using the same initial data.
All results presented here use $r_0 = 0.7$ and $\sigma = 0.3$ and 
a radial resolution of $10000$ gridpoints.
The criticality parameter $p$ is identified with the amplitude $A$. 
Fig. \ref{fig:psi-perspective} shows $|\psi(\tau,x)|$ for a
near-critical evolution with amplitude $A = 0.144930343980315$.

As has become common practice, we find near-critical data through
a bisection search in $p$. This procedure yields in particular
a numerical approximation to the critical value $p=p*$
which defines the threshold of singularity formation.

In the bisection search, a number of criteria are possible to distinguish
dispersion from collapse. Their equivalence can be checked a posteriori 
when comparing the final result, i.e. subcritical and supercritical solutions
close to criticality.
A typical criterion is to monitor the ratio $2 m/r$, where $2 m/r = 1$
signifies the presence of an apparent horizon. This criterion has been
used successfully also in combination with slicing conditions, that do not
penetrate apparent horizons -- as is the case in our approach.
Numerically, we have used the condition $2 m/r \geq 0.995$ as the threshold for
apparent horizon formation and for estimating the black hole mass.
Remarkably, in practice it turns out, that  $2 m/r \geq 0.6$ is a
sufficient criterion to mark a scalar field evolution as supercritical and
thus is useful to speed up bisection searches.
For practical and historical reasons, this is the approach we have adopted 
for our code.

\begin{figure}
\centering
\includegraphics[clip,width=.5\textwidth]{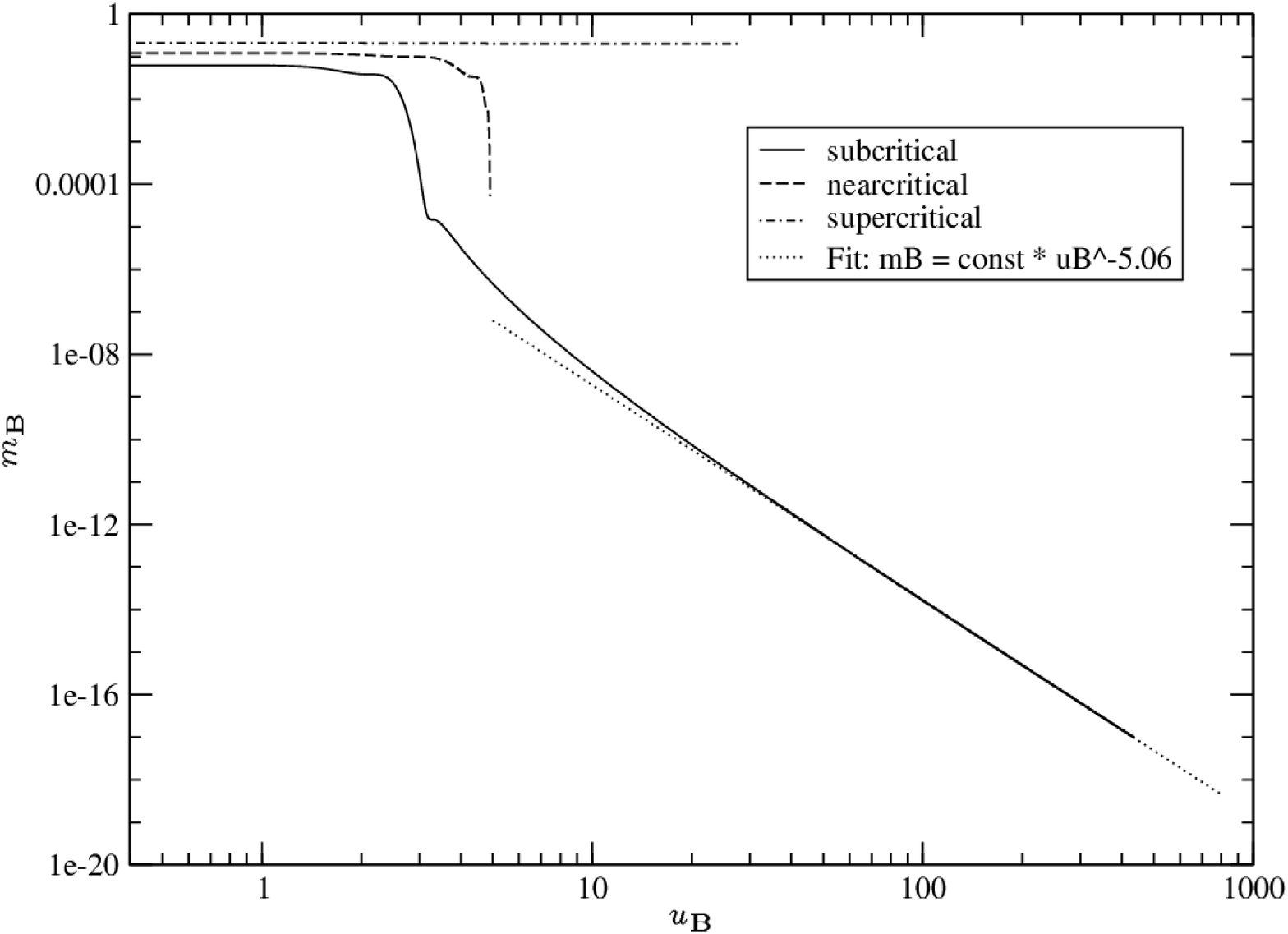}
\caption{\label{fig:m_tail}We compare the decay of the Bondi mass in supercritical, 
near--critical and subcritical evolutions. In the subcritical case, the Bondi mass 
is found to decay for late times with a power-law exponent of approximately $-5$.}
\end{figure}

A number of other options come to mind, in particular in our context of 
evolving out to null infinity, one could e.g. monitor the redshift or Bondi mass.
In a dispersion evolution, the redshift will decay to zero, while it will 
approach infinity when a black hole forms.
Similarly, the Bondi mass will decay to zero with a characteristic 
tail behavior, as is shown in Fig. \ref{fig:m_tail}, in
a dispersion evolution, and will asymptote to a (positive) constant 
when the field does not disperse.

\begin{figure}
\centering
\includegraphics[clip,width=.5\textwidth]{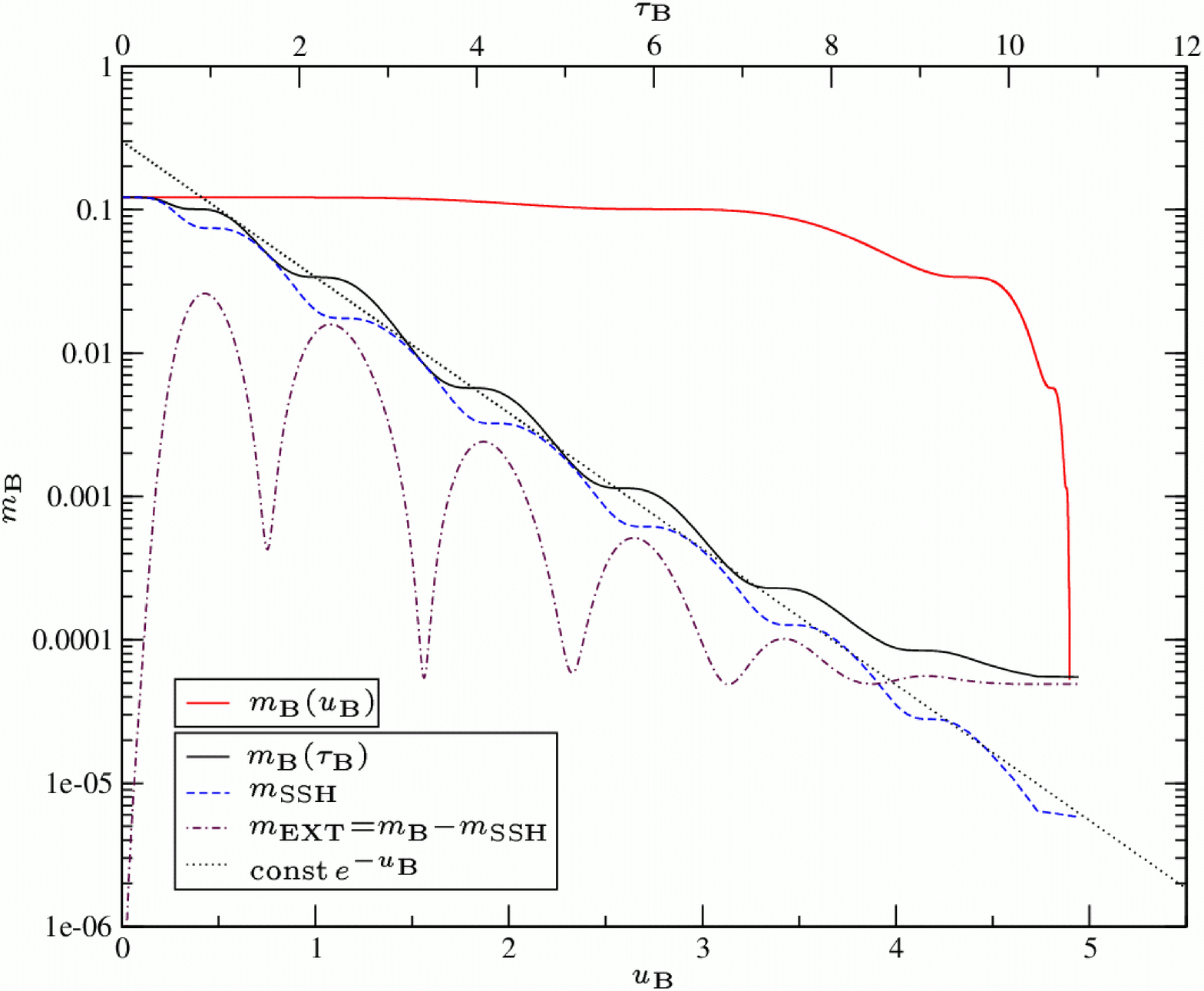}
\caption{\label{fig:mass}This figure plots the Bondi mass $\mB$ against both $\uB$
and the adapted time $\tauB$ for a barely supercritical evolution with final
black hole mass $M_f \approx 5 \times 10^{-6}$. The Bondi mass $\mB$ and the mass at
the past SSH, $\mSSH$, are found to decrease exponentially in $\tauB$ 
(with an overlayed $\tauB$-periodic oscillation with period 
$\Delta /2$), once the evolution has sufficiently approached the critical solution
near the center of spherical symmetry. 
We also show $\mEXT$, the energy present outside of the SSH.
}
\end{figure}

\begin{figure}
\centering
\includegraphics[clip,width=.5\textwidth]{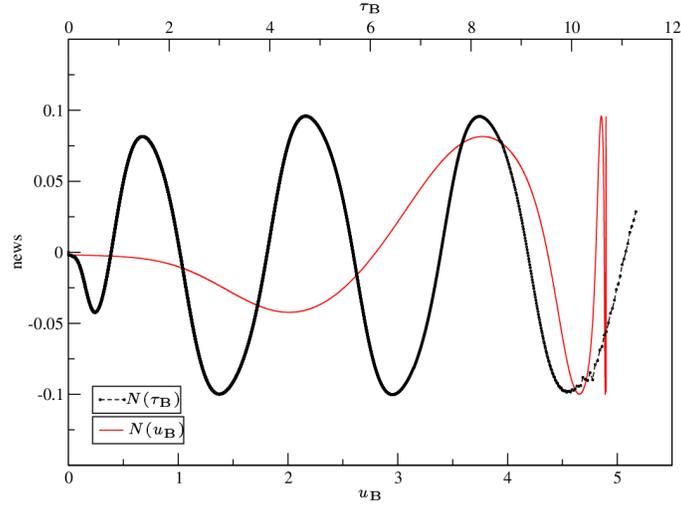}
\caption{\label{fig:news}We show the news function $N(u)$, first as a function of
the natural time coordinate $\uB$ of an asymptotic observer 
and also as a function of a suitably adapted time
$\tauB = -\ln \frac{\uB^* - \uB}{\uB}$ where $N(\tauB)$ is periodic 
with period $\Delta \simeq 3.44$ after the spacetime has come close to the
critical solution.
Even if the constant $\uB^*$ is not known, it can be determined by
a fit to periodicity in $\tauB$. Thus, it is possible to observe
DSS at $\Scri^+$ and to extract the critical exponent $\Delta$.
}
\end{figure}

In the course of a near-critical evolution, 
remnants of the self-similar dynamics 
which occur locally, inside the SSH, are radiated to future null infinty. 
Remarkably, we observe that the imprints of DSS behavior are still present in 
asymptotic quantities such as the Bondi mass and the news function 
(see figures \ref{fig:mass} and \ref{fig:news}).

\begin{figure}
\centering
\psfrag{'masses' using (lp($1)):(log($5) - 0.373*lp($1) +0.74)}[][]{}
\includegraphics[angle=-90,clip,width=.5\textwidth]{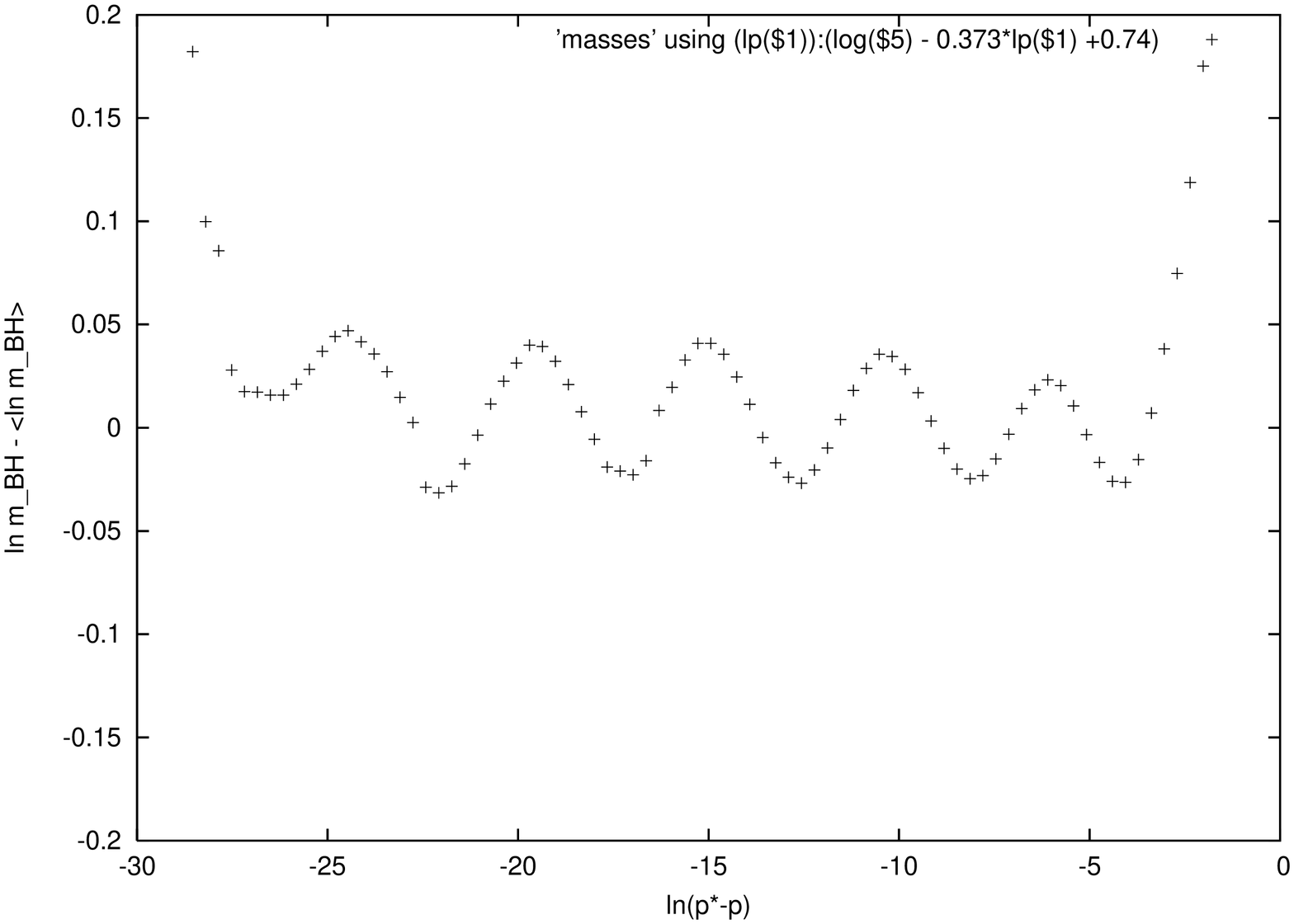}
\caption{\label{fig:fine-structure}
The fine-structure in $\mBH$ after subtracting a linear fit. The measured period $4.6$
is close to the value predicted by perturbation theory $\frac{1}{2}\Delta/\gamma \approx 4.61$.
}
\end{figure}

In addition, in numerical evolutions of supercritical data, the black hole mass
has been found to exhibit a universal scaling law 
(see \cite{Hod97,Koike95,Gundlach97f}):
\begin{equation}
\ln \mBH = \gamma \ln(p-p^*) + \Psi(\ln(p-p^*)) + const,
\end{equation}
where $\gamma \approx 0.373$ and the function $\Psi$ is periodic with 
period $\frac{1}{2}\Delta/\gamma$
in $\ln(p-p^*)$.
One needs to ensure to measure the mass of the outermost component of the 
horizon in order to obtain the fine-structure of the scaling law 
shown in figure \ref{fig:fine-structure}.

This scaling law has been derived by arguments based on linear
perturbation theory around the  critical solution and dimensional analysis
however without using a precise definition of the black hole mass 
\cite{Hod97,Gundlach97f}.
A typical approach in numerical simulations based on coordinates that do not
penetrate apparent horizons, as far as we are aware, is
to follow a peak in $2m/r$ until this quantity almost reaches unity, at which
point the simulation is usually slowed down by a Courant-Friedrichs-Levy (CFL)-type 
condition, and then read off the approximate horizon mass at this point. 
We follow this heuristic approach, and are able to reproduce the mass scaling 
and fine structure.

At least conceptually, this approach is problematic without quantifying how
much mass-energy still remains outside of the peak in $2m/r$ where
the horizon-mass is read off. From the perturbation theory argument,
one expects scaling behavior only for quantities within the self-similar
region of spacetime, which does not necessarily extend beyond the SSH. 
We will return to this point in the discussion section \ref{sec:discussion}.

Note that for a near-critical solution, the approximate value of the
accumulation
time naturally defines an approximate location (i.e. advanced time) of the SSH.

For observers at null infinity, a natural time coordinate is Bondi time
$\uB$ as defined in eq. (\ref{eq:def_BondiTime}) -- Bondi time can be
identified with the proper time of timelike observers
at large distances -- see the discussion in Sec. \ref{sec:discussion}.
In order to analyze critical phenomena, we define
a time coordinate which is suitably adapted to self-similar critical collapse
and set 
\begin{equation}\label{eq:tauB-def}
\tauB = -\ln \frac{\uB^* - \uB}{\uB}.
\end{equation}
The adapted time coordinate $\tauB$ can be defined for spacetimes which are
close to the critical solution inside of the SSH, so that 
the value of the accumulation time $\uB^*$ can be determined by
a fit to periodicity in $\tauB$. We have used a fit to periodicity
of the news function -- $N(\tauB)$ is periodic 
with period $\Delta \simeq 3.44$ after the spacetime has come close to the
critical solution (see figure \ref{fig:news}).

Note that $\tauB$ is only an approximate adapted coordinate since it depends on the relation
between Bondi and central time \eqref{eq:def_BondiTime}.
In order to gain some insight into the behavior of $\tauB(\tau)$ consider
the simpler case of a continuously self-similar (CSS) collapse. We assume that 
$\beta$ changes only little outside the past SSH, i.e. $H(u) \approx \beta_\SSH(u)$.
Then, since $\beta_\SSH(u) = constant$ in the CSS regime, we find by integrating
\eqref{eq:def_BondiTime} that
\begin{equation}
\uB \approx C u,
\end{equation}
where $C = e^{2\beta_\SSH}$ and we have chosen the initial condition $\uB(0) = 0$. 
Furthermore, it follows from the definition of the adapted time coordinates, 
equations \eqref{eq:tau-z-def} and \eqref{eq:tauB-def}, that
\begin{equation}\label{eq:tauB-tau-rel}
\tauB \approx \tau,
\end{equation}
in the CSS regime.
Numerically, the deviations from this relation for the DSS case turn out to be
quite small.

\subsection{DSS behavior in the Bondi Mass and the News Function}\label{sec:mass_news}

\begin{figure}
\centering
\psfrag{Scri+}[][]{$\scriptstyle \Scri^+$}
\psfrag{r=0}[][]{$\scriptstyle z=0$}
\psfrag{past SSH}[][]{\scriptsize past SSH}
\psfrag{z=1}[][]{$\scriptstyle z=1$}
\psfrag{future SSH}[][]{\scriptsize future SSH}
\psfrag{z=infty}[][]{$\scriptstyle z=\infty, \tau=\infty$}
\psfrag{tau=const}[][]{$\scriptstyle\tau = const$}
\includegraphics[clip,width=.3\textwidth]{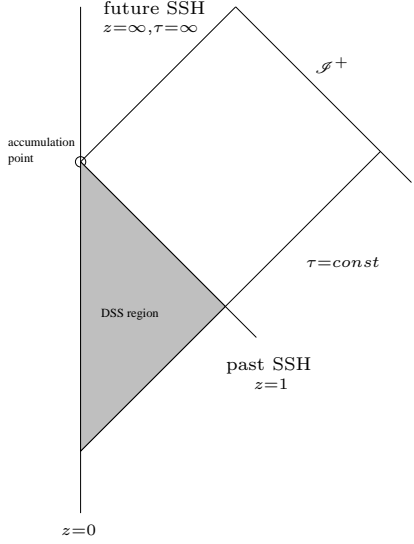}
\caption{\label{fig:SSH}A conformal diagram of a critical collapse spacetime.
In the backwards light cone of the accumulation point the dynamics are close
to the DSS critical solution. The lightlike boundary is also called the
past self-similarity horizon (SSH). Depending on whether the initial data are sub- or
supercritical, the spacetime will for late times be Minkowski or Schwarzschild.}
\end{figure}

The following argument suggests that the news function is approximately 
periodic in $\tauB$, as shown in figure \ref{fig:news}.
Assume effective DSS data for the scalar field and the metric functions on 
the past self-similarity horizon (SSH) (see figure \ref{fig:SSH}). 
Moreover, we assume that the contribution of the right hand side integral in the wave 
equation for the scalar field \eqref{eq:phi-evo-continuum} can be neglected outside of 
the SSH, such that the DSS data on the SSH are linearly propagated to $\Scri^+$ 
without backscattering, i.e. 
\begin{equation}
\lim_{r\to\infty} \psi(u,r) \approx \psi_\SSH (u).
\end{equation}
Furthermore, assume that changes in $\beta$ outside of the SSH are small, so that
$\beta_\SSH(u) \approx H(u)$.
It then follows that
\begin{equation}
\begin{split}
N(\uB) &= \frac{d c(\uB)}{d\uB}  \approx \frac{d\psi_\SSH(u)}{d\uB}\\
       &= \frac{d}{d\tau}\Bigl[ \phi_\SSH(\tau) \zeta(\tau) e^{-\tau} u^* \Bigr] \frac{e^\tau}{u^*} e^{-2H} \\
       &= e^{-2\beta_\SSH} \Bigl[ \frac{d}{d\tau} \left(\phi_\SSH(\tau) \zeta(\tau) \right)
                                - \phi_\SSH(\tau) \zeta(\tau) \Bigr],
\end{split}
\end{equation}
which shows that the news function $N(\uB)$ is approximately
periodic in $\tau$ (and in $\tauB$ if equation \eqref{eq:tauB-tau-rel} holds) 
with period $\Delta$ and satisfies $N(\tau + n\Delta/2) = (-1)^n N(\tau)$.

In order to determine the behavior of the Bondi mass,
we can then rewrite the Bondi mass-loss equation \eqref{eq:mass-loss} 
\begin{equation}
\frac{d\mB}{d\tauB} = -4\pi N^2(\tauB) \frac{d\uB}{d\tauB} = - 4\pi \uB^* e^{-\tauB} N^2(\tauB).
\end{equation}
Since $N^2(\tauB)$ is $\Delta/2$-periodic in $\tauB$, $\mB$ then takes the following form 
\begin{equation}\label{eq:mBondi-DSS}
\mB(\tauB) \approx e^{-\tauB} f_\SSH (\tauB),
\end{equation}
where $f_\SSH (\tauB)$ is $\Delta / 2$-periodic in $\tauB$.

This behavior mimics the behavior of the mass-function:
We rewrite the mass-function in adapted coordinates $(\tau,z)$, using \eqref{eq:z-def},
\begin{equation}
m(\tau,z) = \frac{1}{2} z e^{-\tau} \zeta(\tau) u^* 
            \Bigl[1 - \Vr(z,\tau) e^{-2\beta(z,\tau)} \Bigr],
\end{equation}
and evaluate it at the SSH, which, by a judicious choice of the periodic function $\zeta(\tau)$ 
can be chosen to be at $z=1$ (since the past SSH is a null surface, one needs to ensure that
$\nabla_a z$ becomes null at $z=1$).
We obtain
\begin{equation}
\mSSH (\tau) = e^{-\tau} f^*_{\scriptscriptstyle\text{SSH}}(\tau),
\end{equation}
where $f^*_{\scriptscriptstyle\text{SSH}}(\tau)$ is periodic with period $\Delta/2$.

The periodicity and exponential decay of $\mB$ and $\mSSH$ are confirmed by our 
numerical calculations (see fig. \ref{fig:mass}).
Note that the Bondi mass levels off at roughly
$10^{-4}$ of the mass in the initial configuration, whereas the mass contained
within the backwards light cone $\mSSH$ continues to scale according to the
prediction of critical collapse evolution, equation \eqref{eq:mBondi-DSS}. 
Since the difference, i.e. $\mEXT = \mB - \mSSH$, 
is almost zero at the initial slice, we conjecture that $\mEXT$ 
is the energy due to backscattering in a critical evolution.

Note that $\mEXT$ oscillates with $\tau$, as can be seen by integrating
\begin{equation}
\pd{m}{r} = 2\pi r^2 \bigl(1-\frac{2m}{r}\bigr) \left(\pd{\phi}{r}\right)^2,
\end{equation}
from the SSH to $\Scri^+$.
The biggest contribution to $\mEXT$ will come from the vicinity of the past SSH.

We observe that $\mEXT$ contained in a slice close to horizon formation
is almost constant for different near critical evolutions (down to the
numerical limit of fine tuning).
Therefore, if this $\mEXT$ eventually falls through the horizon, then the resulting 
black hole will have a tiny but finite Bondi mass, no matter how fine tuned the
data are
.
In the very late stages of the evolution, the growing redshift, $\beta \to \infty$ 
effectively halts the numerical evolution, while the error norm 
$\max |\mathbf{E_{uur}}(\tauB > 10)|$ approaches $10^{-1}$.

\subsection{Quasinormal Modes}\label{sec:QNMs}

In gravity, quasinormal modes (see \cite{Kokkotas99a} for a review) are excitations 
of a black hole (or a star) satisfying radiation boundary conditions.
These excitations are, in general, obtained from linear perturbations off a fixed 
background, together with their associated (complex) eigenvalues. 
Thus, in a highly dynamical setting, such as in critical collapse evolutions, 
one would not expect to see (identify) quasinormal modes.
However, it turns out that for our setting, the least damped spherically symmetric mode 
for scalar perturbations of a Schwarzschild black hole plays a relevant role.

Perturbation theory\cite{Iyer87a} gives the following value for the half-period
\begin{equation}
\frac{T_0}{2} \approx 28.43 M_{bg}.
\end{equation}
This mode has previously been detected in supercritical evolutions (far away from criticality) 
for a self-gravitating massless scalar field by Gundlach et al. \cite{Gundlach94b}. 

\begin{figure}
\centering
\includegraphics[clip,width=.5\textwidth]{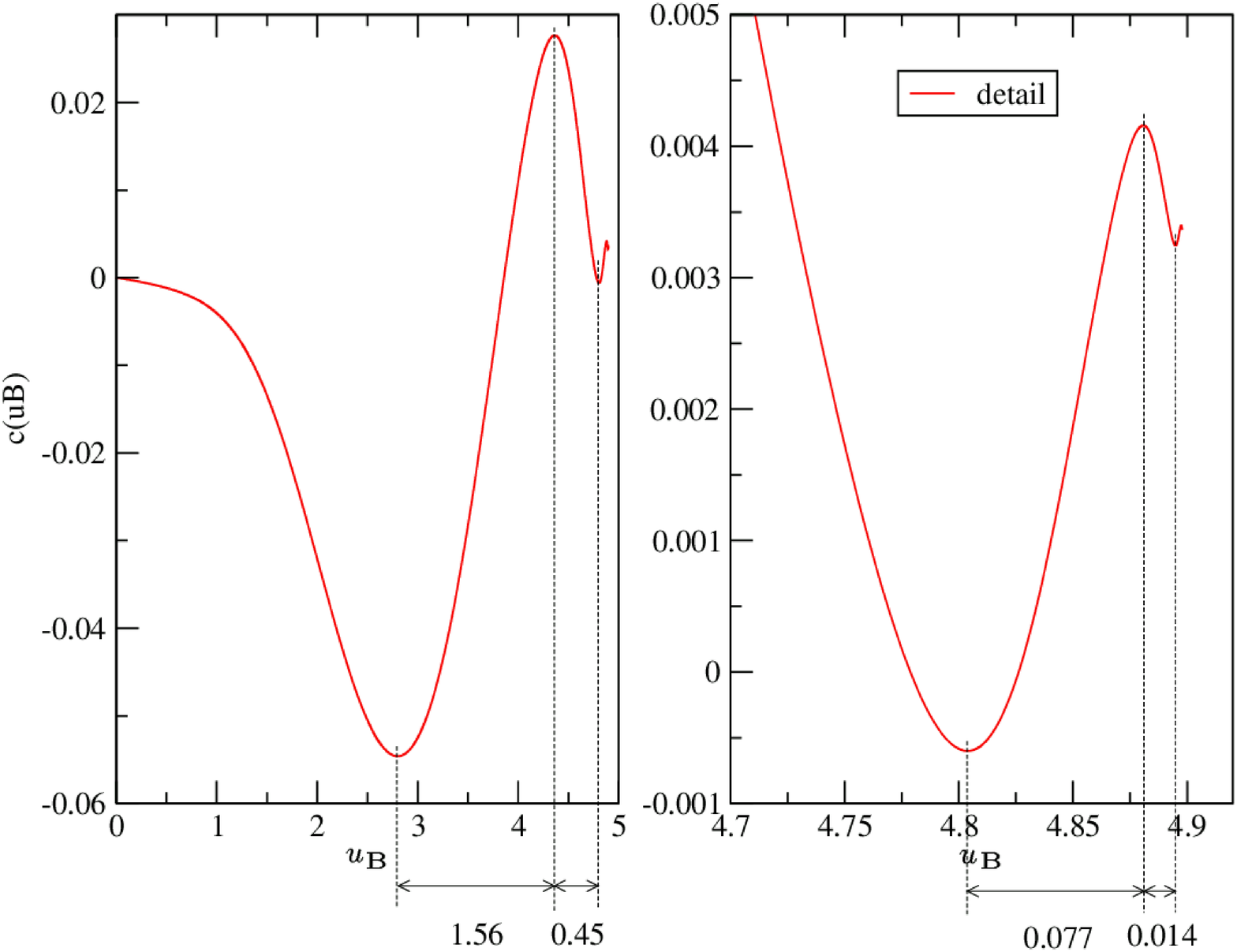}
\caption{\label{c_QNM}This figure shows the scalar field monopole moment $c(\uB)$ for a 
near-critical (but barely supercritical) evolution. The half-periods
measured from one extremum to the next roughly agree with the prediction
of perturbation theory shown in figure \ref{T0_QNM}.
}
\end{figure}

\begin{figure}
\centering
\psfrag{tau_B}[][]{$\scriptstyle \tauB$}
\psfrag{T0/2(u_B)}[][]{$\scriptstyle T_0/2 (\uB)$}
\psfrag{T0/2 = 28.43 * m_B}[][][0.9]{$\scriptstyle T_0/2 \,=\, 28.43\,\mB$}
\includegraphics[clip,width=.5\textwidth]{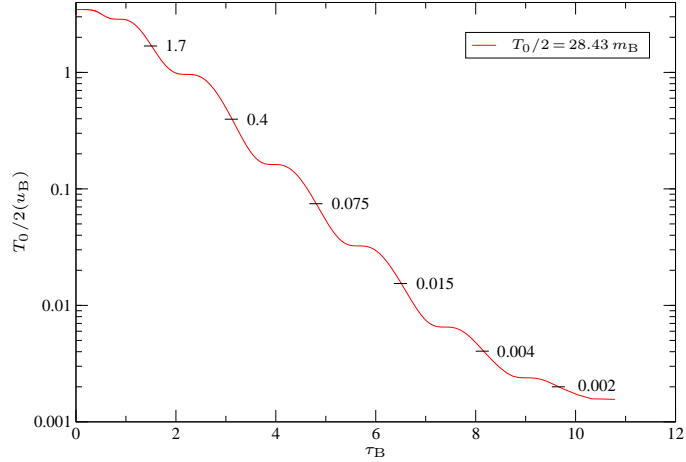}
\caption{\label{T0_QNM}The exponential decay of the QNM half-periods predicted by 
perturbation theory is shown with annotated values at the midpoints
between the points of inflection for the same near--critical evolution as
in figure \ref{c_QNM}. 
}
\end{figure}

In the following, we analyze radiation signals for near-critical evolutions,
where the notion of a fixed background mass does no longer apply.
We find that the monopole moment of the scalar field $c(\uB)$ shows a damped 
oscillation with exponentially decreasing frequency (see figure \ref{c_QNM}).
Moreover, the sizes of the half-periods measured from one extremum to the next in
$c(\uB)$ roughly agree with the half-periods obtained from
the least damped quasi-normal mode (QNM) of a Schwarzschild black hole with a 
strongly changing ``background'' mass $M_{bg}(\uB)$ as shown in figure
\ref{T0_QNM}.
$M_{bg}(\uB)$ is obtained by evaluating $\mB(\uB)$  at the mean value 
between the extrema of $c(\uB)$ (which are inflection points of $\mB(\uB)$).

\subsection{Power-law Tails}\label{sec:tails}

As has been established by Price \cite{Price72}, perturbation fields outside
of Schwarzschild black holes die off with an inverse power-law tail at late times.
In contrast to quasinormal modes this behavior does not depend on the details of the
collapse process, but only on the asymptotic falloff of the effective potential;
(i.e. in a curved spacetime, wave propagation is not confined to the light cones,
rather waves spread inside the light cones, due to scattering off spacetime curvature.)
Therefore, tail phenomena can be observed independently of the endstate of the evolution.
For Gaussian initial data the Newman-Penrose constant \cite{Gomez94b}
vanishes and for late times perturbation theory \cite{Gundlach94a,Gundlach94b}
predicts that the field falls off as 
$\phi \propto \uB^{-2}$ near $\Scri^+$ and $\phi \propto \uB^{-3}$
near timelike infinity $i^+$.

To be more precise \cite{Barack99a,Barack99b}, consider a distant static 
observer at $r=const$ and Bondi time $\Delta \uB$ elapsed since the ``main pulse'' 
of radiation has reached the observer (the duration of the main pulse
has been assumed negligible in \cite{Barack99a,Barack99b}). 
Null infinity is then found to be approximated by the 
region where $\Delta \uB \ll r$ within the context
of a perturbative analysis of tail behavior \cite{LeaverJMP,LeaverPRD,Barack99a}.
This regime has been termed the ``astrophysical zone'' by
Leaver \cite{LeaverJMP,LeaverPRD,Barack99a}.

In addition, the convergence of the perturbation expansion in \cite{Barack99a,Barack99b} 
requires $\Delta \uB \gg M$, where $M$ is the mass of the background. 
In our case, the mass which gives rise to the effective potential is bounded 
from above by the mass of the initial (ingoing) pulse, $M_i$. Therefore, we 
demand that $\Delta \uB \gg M_i$.
Numerically, we observe tails only for $\Delta \uB > 10^3 M_i$.

\begin{figure}
\centering
\psfrag{uB}[][]{$\scriptstyle \uB$}
\includegraphics[clip,width=.6\textwidth]{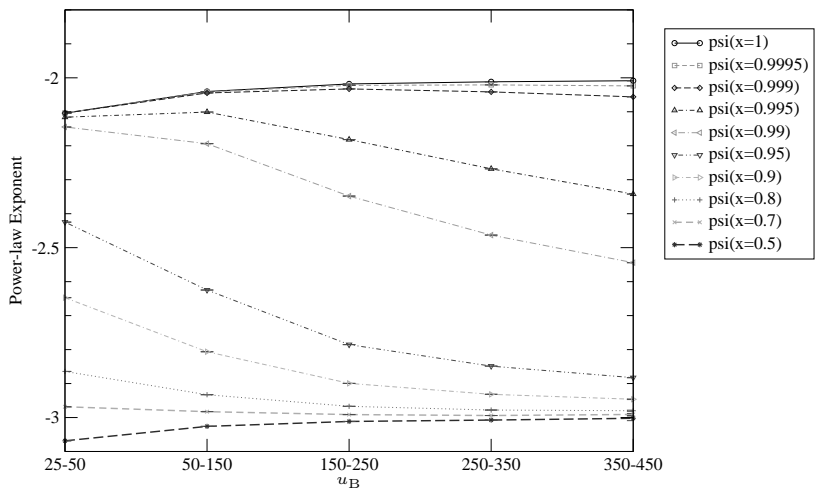}
\caption{\label{fig:power-law}This figure shows power-law exponents for a 
subcritical evolution and illustrates the domains of validity of the 
predictions of perturbation theory for the two zones: $-2$ near 
$\Scri^+$ and $-3$ near $i^+$. 
}
\end{figure}

Closeness to timelike infinity, on the other hand, demands $\Delta \uB \gg |r_*|$,
where $r*$ is the usual ``tortoise'' coordinate $r_* = r + 2M \ln \bigl( \frac{r-2M}{2M} \bigr)$.
In figure \ref{fig:power-law} we show power-law exponents determined
by fits of $\psi$ at different $r=const$ curves over a series of time intervals for a 
subcritical evolution.
As described in section \ref{sec:algorithm} we use spline interpolation to 
calculate $\psi(x=const)$.
The exponents have been determined by fitting the field $\psi$ at $x=constant$ 
against a power of $\uB$ in 5 distinct time intervals of the evolution.
The domains of validity of the exponents predicted by perturbation theory, $-2$ near 
$\Scri^+$ and $-3$ near $i^+$, can be observed here. 
It is clear that the outermost gridpoints in this evolution (using 10000 gridpoints) 
are indeed located in the ``astrophysical zone'' since $r(x=0.9995) \approx 2000 \gg \Delta \uB$,
where $\Delta \uB$ is the Bondi time elapsed since the main pulse of radiation has reached
the observer at about $\uB \approx 3$ (see figure \ref{fig:m_tail}). 
On the other hand, closeness to timelike infinity $i^+$ demands that
$\Delta \uB \gg r_*$. Note that $r_* \approx r$ for $r \gg 2M_i$, where $M_i \approx 0.06$ 
is the initial Bondi mass. 
It is also apparent that for $\Delta \uB \approx r$ the observers are in between the two zones
and the power-law exponents seem to change smoothly.

Figure \ref{fig:m_tail} displays the power-law decay, with exponent $-5$, of the Bondi mass
for the same subcritical evolution.
This behavior can be explained by integrating the Bondi mass-loss equation \eqref{eq:mass-loss}
with $c \propto \uB^{-2}$ in the regime of power-law tails.

Note that in situations involving realistic sources and realistic detectors, 
power-law tails only play a minor (if any) role as an actual signal, but the power-law
tail results still can provide hints on how to resolve the very important question
of whether null infinity is a useful idealisation for gravitational wave detectors.
Accordingly, we suggest to generalize the term ``astrophysical zone'' to
more general, non-perturbative situations, by reinterpreting -- in a very loose sense --
$\Delta \uB$ as a suitably chosen large time scale characteristic of the source.
The physical idea is that the distance from observers of astrophysical phenomena, 
e.g. gravitational wave detectors, to the radiation sources is very large compared 
to the time during which substantial radiation from the source can be observed.
This is at least expected for sources where general
relativity is important, as opposed to problems for which a (Post-) Newtonian approach
and the quadrupole formula are sufficient. An example would be a binary black hole  
merger in another galaxy, which might
have a characteristic dynamical time scale of a fraction of a second, and which 
might be observed during several thousand cycles, including a portion of the QNM 
ringdown.

\section{Discussion}\label{sec:discussion}

In this work we have presented numerical constructions of portions of
near-critical spherically symmetric spacetimes that extend up to future null
infinity and asymptote to the event horizon.
The simulations are based on a compactified approach, where the equations
have been regularized in a neighborhood of null infinity by introducing the
mass function as an additional independent variable.
The resolution necessary to resolve critical phenomena is gained through
letting our gridpoints fall along ingoing null geodesics, following Garfinkle
\cite{Garfinkle95}. The grid is repopulated with grid points, which are filled
with values through interpolations when half of the gridpoints have reached
the origin.
This is not optimal, 
but ensures sufficient resolution for our current purposes.

We reproduce standard features of near-critical solutions, such as 
``echoing'' and mass scaling with fine structure. In addition
we extract radiation at null infinity by computing the news function and find
a signal with rapidly increasing frequency as measured in Bondi time, which is
the natural time coordinate associated with far away observers. In order to
simplify the analysis of near-critical spacetimes, we
have defined an approximate adapted time in terms of Bondi time, in analogy
to the standard time coordinate which is adapted to the DSS discrete 
diffeomorphism (\ref{DSS-def}).
This coordinate can be used by asymptotic observers to render the signal
from a near-critical collapse (almost) periodic.
The fact that such a definition actually works out, and makes DSS periodicity
manifest in quantities defined at null infinity is a non-trivial result, which
we explain in Sec. \ref{sec:crit_behavior}.

Note also that the amplitude of the news function, shown in Fig. 
\ref{fig:news} stays fairly constant after the initial transient.
This feature is clearly universal, as long as the radiation signal is dominated
by the DSS collapse, since the system can be approximated
in the DSS region by a perturbation of the critical solution,
and, according to our results, the radiation signal is dominated by
the DSS structure of the critical solution. Thus, in section \ref{sec:mass_news},
we neglect contributions from scalar field which is far outside the DSS region 
and does not contribute to the critical collapse dynamics.
Consequently, in this scenario the essential free parameters 
determining the radiation signal from an actual near critical solution
are the number of cycles the
solution spends in the neighborhood of the critical solution and
the length scale at which the solution comes close to the critical
solution. The robustness of this scenario, i.e. what happens if
the initial data are such that there is significant mass outside of
the SSH, is outside of the scope of this paper and an issue for future research.

Perhaps the most surprising feature of the radiation signal has emerged
from our investigation of QNM's, which has been motivated by
\cite{Gundlach94b}, where  the first quasinormal mode is found in collapse
evolutions and the question is posed as to how QN ringing would change close
to criticality.
We find that even in very close-to-critical evolutions there is
a correlation of the radiation signal with the period of the first quasinormal 
mode, determined from the time-dependent value of the Bondi mass, as disussed 
in Sec. \ref{sec:QNMs}, Figs. \ref{c_QNM} and \ref{T0_QNM}. 
This correlation between the radiation of the highly dynamical near-critical 
solution and the quasinormal mode, that is defined in terms of perturbations of
a static spacetime, certainly deserves further investigation. 
This surprising feature might even turn out to be a key toward understanding 
the phenomenon of DSS behaviour in near-critical spacetimes.
Our results seem to suggest that the effective curvature potential for a DSS
self-gravitating field acts as a quasi-stationary background for
scattering processes which can be approximately described by quasinormal modes
of a 1-parameter family of Schwarzschild black holes with exponentially
decreasing mass. The question of the applicability of QNM-motivated estimates
is quite relevant for numerical relativity, e.g. when extracting wave
forms from binary mergers.
In the very different context of quasinormal modes of Schwarzschild-AdS black
holes, Horowitz and Hubeny \cite{Horowitz00} have pointed out an agreement of
the numerical values of the critical exponent and the imaginary part of
quasinormal frequencies, and have speculated on a connection between
critical collapse and quasinormal modes. While Horowitz and Hubeny assume
pure coincidence as the most likely explanation, it seems worth to keep in
mind in future work concerning critical collapse or quasinormal modes.

For subcritical initial data we can evolve for very long times, and thus are
able to observe power-law tail behavior as shown in Figs. \ref{fig:power-law}
and \ref{fig:m_tail}.
Analytical calculations predict different falloff rates for radiation along
null infinity and along timelike lines, and the natural question arises, which
falloff rate would be seen by a hypothetical observer (in a realistic case,
observation of power-law tails would require an extremely large signal-to-noise 
ratio).
Accordingly, our results depicted in Fig. \ref{fig:power-law} show how
the rates at finite but large radius correspond to the value for null infinity
for a while before they approach the expected late-time value for finite
radius.
The interpretation of this phenomenon is suggested by
perturbative work of \cite{Barack99a,Barack99b}, 
where different tail falloff rates are computed.
There, different regions of spacetime are identified, where certain 
approximations hold. Within the perturbative regime, results obtained 
for null infinity are valid for what has been termed the
``astrophysical zone'' by Leaver \cite{LeaverJMP,LeaverPRD,Barack99a}, 
and which is defined as the region where $\Delta \uB \ll r$.
(Note, however, that $\uB \gg M$ must also be satisfied).
The physical idea is that the distance from observers of astrophysical
phenomena, e.g. gravitational wave detectors, to the radiation sources is 
very large compared to the time during which the source radiates at an observable rate.

We argue that our results illustrate that the relevant falloff from the point
of view of an astrophysical observer is the falloff rate at null infinity,
in accordance with the prediction from perturbation theory. 
We believe that this is a nice model calculation that exemplifies that
null infinity is indeed a useful approximation for observers at large distance
from the source, in the sense that such observers are located in the 
``astrophysical zone''.

Along the same lines, we would like to point out that by
taking appropriate limits in a conformally compactified manifold,
worldlines of increasingly distant geodesic observers converge to null
geodesic generators of future null infinity
and proper time converges to Bondi time \cite{Frauendiener98c}.
Note also that a naive correspondence between observers at large distance 
and spatial infinity would be problematic, e.g. compactification at spatial
infinity leads to ``piling up'' of waves, whereas at null infinity this effect
does not appear -- waves leave
the physical spacetime through the boundary at null infinity.

Under practical circumstances, e.g. computing the
signal from a source of radiation which is locted at a cosmological
distance from the detector, null infinity more realistically
corresponds to an observer that is sufficiently far away from the source
to treat the radiation linearly, but not so far away that cosmological
effects have to be taken into account. We are however not aware of
a discussion where this sloppy picture has been made more precise.

When looking at critical collapse from a global spacetime perspective,
as we have done here, one is confronted with some issues concerning
the mass-scaling, that we would like to comment on briefly:
When asking the question of whether infinitesimally small black holes can be
 formed 
-- the question which triggered the original work on critical collapse --
it could be phrased in two slightly different ways: (i) can we
form black-hole solutions where the final-state black hole has arbitrarily
small mass, or (ii) can we form arbitrarily small apparent horizons.
The question which has been aswered in the affirmative by critical collapse
research is the second one. The first one still seems open. 
Our results for near-supercritical evolutions seem to indicate that the
final black hole is significantly larger than the mass leading to scaling.
We conjecture that due to backscattering of the outgoing radiation one
cannot form arbitrarily small black holes, no matter how fine tuned the data are.

Finally, we want to emphasize the obvious fact that the failure of the
critical solution to be asymptotically flat is perfectly consistent with its
role in the dynamics of a localized object emitting radiation to null infinity.
First, note that for the dynamics of critical collapse, only a small region of
spacetime is relevant. The scale-invariance of the DSS solution is compatible
with asymptotic flatness in the following way: a near critical solution, which
may or may not be asymptotically flat, comes close 
to the critical solution at a length scale which depends on the initial data,
follows the critical solution for a number of cycles, and then collapses
or disperses. The (discrete) self-similarity of the critical solution
is on the one hand responsible for the failure of the critical solution to be
asymptotically flat but actually makes it possible to keep a free length
scale in the problem of massless scalar field collapse, as is expected to
allow for asymptotically flat near critical solutions.

\begin{acknowledgments}
This work has been supported in part by the Austrian Fonds zur F\"orderung
der wissenschaftlichen Forschung (FWF) (projects P12754-PHY and P15738-PHY).
S.H. has been supported in part by the grant BFM2001-0988 sponsored by 
the Spanish {\em Ministerio de Ciencia y Tecnolog\'ia}.
We thank Christiane Lechner and Jonathan Thornburg for the contributions
to the numerical evolution code \cite{Husa2000b} on which the code used 
here is based. 
S.H. acknowledges the hospitality at the University of Vienna,
and the University of Pittsburgh in the early stages of this work.
M.P. and P.C.A. thank the Albert-Einstein-Insitut in Potsdam for hospitality.
P.C.A. acknowledges partial support by the Fundacion Federico.
M.P. also thanks Jos\'e M. Mart\'\i n-Garc\'\i a for stimulating discussions.

\end{acknowledgments}


\bibliographystyle{apsrev}
\bibliography{./bibtex/references}

\end{document}